\documentclass[twocolumn,english,aps,prb]{revtex4-1}
\usepackage[latin9]{inputenc}
\usepackage[active]{srcltx}
\usepackage{amsmath}
\usepackage{amssymb}
\usepackage{graphicx}
\usepackage{esint}

\makeatletter
\@ifundefined{textcolor}{}
{%
 \definecolor{BLACK}{gray}{0}
 \definecolor{WHITE}{gray}{1}
 \definecolor{RED}{rgb}{1,0,0}
 \definecolor{GREEN}{rgb}{0,1,0}
 \definecolor{BLUE}{rgb}{0,0,1}
 \definecolor{CYAN}{cmyk}{1,0,0,0}
 \definecolor{MAGENTA}{cmyk}{0,1,0,0}
 \definecolor{YELLOW}{cmyk}{0,0,1,0}
}

\usepackage{amsfonts}%\usepackage{dsfont}
\usepackage{mathrsfs}
\usepackage{bbm}
\usepackage{bbold}
\usepackage{subfigure}%p\usepackage{caption}
\usepackage{epstopdf}

\newcommand{\rmd}{{\rm d}}
\usepackage{babel}

\usepackage{babel}

\makeatother

\usepackage{babel}
\begin{document}

\title{Phenomenological theory of the superconducting state inside the hidden-order
phase of URu$_{2}$Si$_{2}$ }

\author{Jian Kang}

\affiliation{School of Physics and Astronomy, University of Minnesota, Minneapolis,
MN 55455, USA}

\author{Rafael M. Fernandes}

\affiliation{School of Physics and Astronomy, University of Minnesota, Minneapolis,
MN 55455, USA}
\begin{abstract}
Recent experiments have unveiled important properties of the ground
state of the elusive heavy fermion $\mathrm{URu_{2}Si_{2}}$. While
tetragonal symmetry-breaking was reported below the hidden-order (HO)
transition at $T_{HO}\approx17.5$ K, time-reversal symmetry-breaking
was observed below the superconducting transition temperature $T_{c}<T_{HO}$.
Although the latter results have been used to argue in favor of a
chiral $d+id$ superconducting state, such an order parameter is incompatible
with broken tetragonal symmetry. Here, we employ a phenomenological
model to investigate the properties of a chiral superconducting state
that develops inside the hidden-order phase. In this case, there are
actually two superconducting transition temperatures: while $T_{c}$
marks a normal-state to superconducting transition, $T_{c}^{*}<T_{c}$
signals a superconducting-to-superconducting transition in which time-reversal
symmetry is broken. In the phase $T_{c}^{*}<T<T_{c}$, the low-energy
density of states $\rho\left(\omega\right)$ is enhanced due to the
crossing of two nodal lines, giving rise to an unusual $\omega\log(\omega)$
dependence of $\rho\left(\omega\right)$, which is manifested in several
thermodynamic properties. We also investigate the emergence of a soft
amplitude gap mode near $T_{c}^{*}$. In contrast to the usual amplitude
mode near a regular normal-state to superconducting transition, this
mode becomes soft near a superconducting-to-superconducting transition,
which in principle allows for its detection by Raman spectroscopy.
Finally, we investigate the impact of twin domains on the anisotropic
properties of the superconducting state, and propose experiments in
mechanically strained samples to explore the interplay between hidden
order and superconductivity in $\mathrm{URu_{2}Si_{2}}$.
\end{abstract}
\maketitle

\section{Introduction}

The nature of the ground state of the body-centered tetragonal compound
$\mathrm{URu_{2}Si_{2}}$ remains one of the most elusive problems
in heavy fermion physics. At the HO transition temperature $T_{HO}\approx17.5$
K, the system displays a sharp specific heat anomaly characteristic
of a second-order phase transition \cite{Palstra85,Maple86,Schlabitz86}.
However, despite nearly thirty years of intense research, the broken
symmetries of this hidden-order phase remain a widely debated topic.
On the theory front, many different order parameters have been proposed
to explain the hidden-order state, from high-rank multiple orders
\cite{Nieuwenhuys87,Barzykin95,Harima10,Ohkawa99,Hanzawa05,Hanzawa07,Haule09,Cricchio09,Kusunose11,Gorkov91,Gorkov92,Ramirez92,Santini94}
to exotic states involving the hybridization of localized and itinerant
states \cite{Mineev05,Ikeda98,Kotetes10,Kotetes14,Varma06,Silhanek06,Chandra02,Tripathi05,Chandra13,Fazekas05,Ikeda12,Balatsky11,Fujimoto11,Ferraz11,Kee12,Mydosh09,Das14}.
On the experimental front, recent measurements have provided important
pieces for this long-standing puzzle. For instance, torque magnetometry
\cite{Okazaki11,Shibauchi12}, x-ray diffraction \cite{Tonegawa14},
and elasto-resistance measurements \cite{IFisherNat15} have reported
a tetragonal ($C_{4}$) to orthorhombic ($C_{2}$) transition simultaneously
to $T_{HO}$, manifested by the inequivalence between the $\begin{bmatrix}1 & 1 & 0\end{bmatrix}$
and $\begin{bmatrix}1 & -1 & 0\end{bmatrix}$ directions of the crystallographic
unit cell (see Fig.~\ref{Fig:Structure}). Whether this is the only
broken symmetry in the hidden-order state -- in which case it would
be classified as a nematic state -- remains an open issue \cite{Sacuto14,Blumberg15}.
For instance, quantum oscillation measurements also suggest a translational
symmetry-breaking along the $c$-axis \cite{Hassinger10}, and neutron
elastic scattering experiments favor a rank 5 multipole order in the
system \cite{Ressouche12}.

In comparison to the hidden-order phase, the superconducting state
of $\mathrm{URu_{2}Si_{2}}$ that appears at $T_{c}\approx1.5$K has
received rather less attention. Understanding its nature is relevant
not only within the bigger picture of superconductivity in heavy-fermion
compounds, but also as a potential tool to probe the properties of
the HO state \cite{Zhu09,Chakravarty}, since superconductivity develops
well below $T_{HO}$ and disappears when the HO phase is suppressed
by external pressure \cite{Uemura05,Jeffries08}. An interesting proposal
based on recent angle-resolved specific heat \cite{Yano08} and thermal
conductivity data \cite{Kasahara09}, which reported indirect evidence
for point and line nodes, is of a chiral $d$-wave superconducting
(SC) state described by the order parameter \cite{Kasahara09}:

\begin{equation}
\Delta(k)=\Delta_{0}\sin\frac{k_{z}}{2}\left(\sin\frac{k_{x}+k_{y}}{2}\pm i\sin\frac{k_{x}-k_{y}}{2}\right)\ .\label{Eqn:SCForm}
\end{equation}

Such a SC state breaks time-reversal symmetry, which seems to be in
agreement with magnetic susceptibility \cite{Li13} and recent Kerr
effect measurements \cite{Schemm14}. However, this order parameter
manifestly preserves $C_{4}$ symmetry, whereas the HO state, from
which SC develops, breaks tetragonal symmetry.

In this paper, we use a phenomenological model to reconcile the proposal
of a chiral $d$-wave state with the experimental observations of
a HO state that breaks $C_{4}$ symmetry. While a phenomenological
approach leaves aside the issue of the microscopic mechanisms involved
in the formation of these phases, it allows for general conclusions
to be drawn regardless of one's favorite order parameter for the HO
phase -- as long as it accounts for $C_{4}$ symmetry breaking. As
a result, it provides general benchmarks that must be satisfied if
indeed the HO state breaks $C_{4}$ symmetry and the SC state is chiral.
One obvious consequence from the fact that SC develops in a $C_{2}$
background, as pointed out in different scenarios \cite{Matsuda14,Volovik93,Sigrist87},
is that the chiral superconducting transition is actually split into
two superconducting transitions $T_{c}$ and $T_{c}^{*}$. While at
$T_{c}$ the system first becomes superconducting, at $T_{c}^{*}<T_{c}$
time-reversal symmetry (TRS) is broken, signaling a SC-SC transition.
Our focus in this paper is on the thermodynamic properties of these
two distinct superconducting phases. We find that, while in the regime
$T<T_{c}^{*}$ the nodal quasi-particle density of states $\rho\left(\omega\right)$
depends linearly on $\omega$ for low-energies, reflecting the presence
of point and line nodes, in the regime $T_{c}^{*}<T<T_{c}$ the density
of states acquires an unusual log-dependence $\rho\left(\omega\right)\sim\omega\log\omega$
due to the crossing between two nodal lines. Such a behavior leaves
signatures in several thermodynamic quantities, such as the specific
heat and the penetration depth.

Furthermore, we investigate how the anisotropies of the SC state in
both regimes -- namely $T<T_{c}^{*}$ and $T_{c}^{*}<T<T_{c}$ --
are affected by the presence of twin domains with different $C_{2}$
orientational order. By calculating the angular dependence of the
specific heat in the presence of a magnetic field, we find that a
twinned sample and a sample with no $C_{4}$ broken symmetry would
display nearly identical behaviors. This helps to reconcile some of
the experimental results that led to the proposal of a chiral SC state,
as in Eq. (\ref{Eqn:SCForm}), with the experimental results that
found tetragonal symmetry-breaking in the HO phase. We propose experiments
in mechanically detwinned samples to unveil the intrinsic anisotropies
of the SC state.

We also find an unusual behavior for the collective SC modes near
$T_{c}^{*}$. In a common normal-state to SC transition, the soft
amplitude SC gap-mode falls into the continuum and is strongly damped.
However, because $T_{c}^{*}$ is a SC-SC transition, the corresponding
soft amplitude mode develops in the background of a superconducting
quasi-particle spectrum. Despite the presence of nodal quasi-particles
in this spectrum, which promote under-damping of the soft amplitude
mode, we argue that this collective mode may still be observed in
the excitation spectrum, as measured by Raman scattering.

The paper is organized as follows: in Section II we present our phenomenological
model for the SC state inside the HO phase and discuss its nodal quasi-particle
spectrum for different temperature regimes. In Section III we study
the impact of twin domains on the anisotropic thermodynamic properties
of the SC state, with particular emphasis on the angle-dependent specific
heat. Section IV is devoted to the investigation of the collective
modes of the SC state. Concluding remarks follow in Section V.

\section{Phenomenological model}

\subsection{Superconducting free energy}

The SC order parameter in Eq. (\ref{Eqn:SCForm}) belongs to the $E_{g}$
irreducible representation of the tetragonal point group. Its degeneracy
stems from the two-dimensionality of the $E_{g}$ representation.
However, because the HO state breaks the $C_{4}$ symmetry, the SC
order parameter must be modified to reflect the new orthorhombic point
group symmetry of the system. Using as a starting point Eq. (\ref{Eqn:SCForm}),
we can describe the chiral $d$-wave SC order parameter inside the
orthorhombic HO phase as a two-component order parameter:

\begin{eqnarray}
\Delta(k) & = & \Delta_{1}\sin\frac{k_{z}}{2}\sin\frac{k_{x}+k_{y}}{2}\nonumber \\
 &  & +e^{i\theta}\Delta_{2}\sin\frac{k_{z}}{2}\sin\frac{k_{x}-k_{y}}{2}\ ,
\end{eqnarray}
where $\theta$ is the phase difference between the two components
of the order parameter, and $\Delta_{i}$ are real order parameters.
TRS is kept intact only if $\theta=0,\pi$. The absolute value of
the gap function is then:

\begin{align}
\left|\Delta\right|^{2} & =\sin^{2}\frac{k_{z}}{2}\left[\Delta_{1}^{2}\sin^{2}\frac{k_{x}+k_{y}}{2}+\Delta_{2}^{2}\sin^{2}\frac{k_{x}-k_{y}}{2}\right.\nonumber \\
 & \left.+2\Delta_{1}\Delta_{2}\cos\theta\sin\frac{k_{x}+k_{y}}{2}\sin\frac{k_{x}-k_{y}}{2}\right]
\end{align}

Thus, tetragonal symmetry requires $\theta=\pm\pi/2$ and $\Delta_{1}=\Delta_{2}$,
but in the HO phase this is not the case.

\begin{figure}[htbp]
\centering \includegraphics[width=0.8\columnwidth]{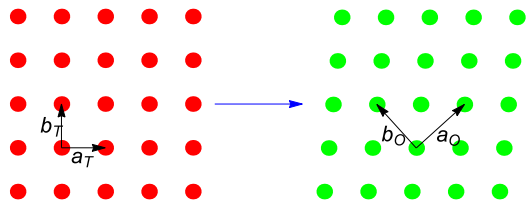} \protect\protect\protect\caption{The basal plane across the HO phase transition. Left: above the HO
phase, the crystal structure (here centered at the U atoms) is body
centered tetragonal. Right: The HO phase breaks the fourfold rotational
symmetry on the $ab$ plane by making the $\begin{bmatrix}1 & 1 & 0\end{bmatrix}$
and $\begin{bmatrix}1 & -1 & 0\end{bmatrix}$ directions (parallel
to the Ru nearest neighbor directions) inequivalent. The crystal structure
becomes base centered orthorhombic.}

\label{Fig:Structure}
\end{figure}

To proceed, we need to establish how the tetragonal symmetry is broken
in the hidden order phase. There are two possibilities: either the
horizontal $\begin{bmatrix}1 & 0 & 0\end{bmatrix}$ and vertical $\begin{bmatrix}0 & 1 & 0\end{bmatrix}$
directions become inequivalent, in which case the $C_{2}$ order parameter
belongs to the $B_{1g}$ irreducible representation, or the diagonal
$\begin{bmatrix}1 & 1 & 0\end{bmatrix}$ and anti-diagonal $\begin{bmatrix}1 & -1 & 0\end{bmatrix}$
directions become inequivalent, in which case the $C_{2}$ order parameter
belongs to the $B_{2g}$ irreducible representation. Torque magnetometry
\cite{Shibauchi12}, x-ray diffraction \cite{Tonegawa14}, and elasto-resistance
measurements \cite{IFisherNat15} all agree on the second scenario,
which is schematically shown in Fig.~\ref{Fig:Structure}.

Thus, we can now write down the phenomenological Ginzburg-Landau model for the superconducting degrees of freedom inside the HO phase \cite{Yanase}:
\begin{eqnarray}
F_{\mathrm{SC}} & = & \frac{a}{2}\left(\Delta_{1}^{2}+\Delta_{2}^{2}\right)+\frac{u}{4}\left(\Delta_{1}^{4}+\Delta_{2}^{4}\right)\nonumber \\
 &  & +\frac{1}{2}\Delta_{1}^{2}\Delta_{2}^{2}\left(\beta+\alpha\cos2\theta\right)-\frac{\eta}{2}\left(\Delta_{1}^{2}-\Delta_{2}^{2}\right)\ ,\label{Eqn:Free}
\end{eqnarray}

Here, $a=a_{0}\left(T-T_{c,0}\right)$ and the order parameter $\eta\neq0$
describes the $C_{4}$ symmetry-breaking inside the HO state. We refrain
from calling it the HO order parameter, since it is unclear whether
other symmetries are also broken in the HO phase. In any case, because
$T_{HO}\gg T_{c}$, we consider $\eta$ to be constant on the temperature
regime in which an expansion in powers of the superconducting order
parameter is allowed. Without loss of generality, we assume in this
section that $\eta>0$ -- in the next section, where we consider the
effects of domains, this assumption is no longer valid.

For the SC state to be a chiral $d$-wave, $\alpha$ must be positive;
furthermore, in order for both order parameter components to coexist,
and for the free energy to be bounded, we must also have $u>\left|\beta-\alpha\right|$.
With this constraints, minimization of the free energy with respect
to $\theta$ gives always $\theta=\pm\pi/2$, as long as both $\Delta_{1}$
and $\Delta_{2}$ are simultaneously non-zero.

\begin{figure}[htbp]
\centering \includegraphics[scale=0.6]{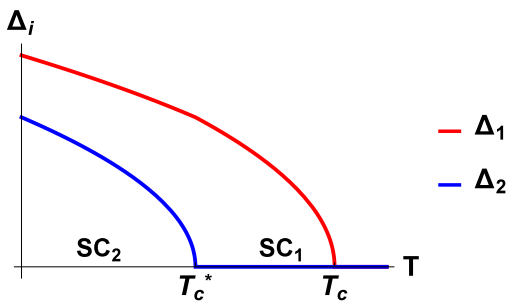} \protect\protect\protect\caption{The SC order parameters in different phases. $\Delta_{1}$ becomes
nonzero when $T<T_{c}$ (SC$_{1}$ phase), and $\Delta_{2}$ becomes
nonzero at $T<T_{c}^{*}$ (SC$_{2}$ phase). The system, therefore,
has two different SC phases. The phase difference between $\Delta_{1}$
and $\Delta_{2}$ is $\pi/2$, i.e. time reversal symmetry is broken
only in the SC$_{2}$ phase.}

\label{Fig:SCOP}
\end{figure}

Minimization of the free energy with respect to $\Delta_{1}$ and
$\Delta_{2}$ reveal two distinct regimes as shown in Fig~\ref{Fig:SCOP},
defined by the two transition temperatures:

\begin{align}
T_{c} & =T_{c,0}+\frac{\eta}{a_{0}}\nonumber \\
T_{c}^{*} & =T_{c}-\frac{2u\eta}{a_{0}\left(u-\beta+\alpha\right)}\label{Tc}
\end{align}

Here, we assume that $a=a_{0}(T-T_{c,0})$ when the temperature is
close to $T_{c,0}$. For $T_{c}^{*}<T<T_{c}$ (SC$_{1}$ phase), the
free energy is minimized by enforcing only one of the SC components
to be non-zero:

\begin{align}
\Delta_{1}^{2} & =-\left(\frac{a-\eta}{u}\right)\:; & \Delta_{2} & =0\nonumber \\
F_{\mathrm{min}} & =-\frac{(a-\eta)^{2}}{4u}\ .\label{SC1}
\end{align}

On the other hand, for $T<T_{c}^{*}$ (SC$_{2}$ phase), the minimum
of the free energy corresponds to the condensation of the two components
with a $\pi/2$ phase difference:
\begin{eqnarray}
\theta & = & \pi/2\ ,\nonumber \\
\Delta_{1}^{2} & = & -\frac{a}{u+\beta-\alpha}+\frac{\eta}{u-\beta+\alpha}\ ,\nonumber \\
\Delta_{2}^{2} & = & -\frac{a}{u+\beta-\alpha}-\frac{\eta}{u-\beta+\alpha}\ ,\nonumber \\
F_{\mathrm{min}} & = & -\frac{a^{2}}{2\left(u+\beta-\alpha\right)}-\frac{\eta^{2}}{2\left(u-\beta+\alpha\right)}\ .\label{SC2}
\end{eqnarray}

Because the free energy changes smoothly, both transitions are second-order.
At $T_{c}$, the system becomes a single-component SC, whereas at
$T_{c}^{*}$, TRS is broken, and the system becomes a two-component
SC. The anisotropy induced in $\Delta_{1}$ and $\Delta_{2}$ is proportional
to $\eta$, as expected, since this order parameter manifestly breaks
the tetragonal symmetry. The splitting between $T_{c}$ and $T_{c}^{*}$
is also proportional to $\left|\eta\right|$.

It is important to discuss the relationship between the orthorhombic
distortion $\delta$ in the $C_{2}$ HO phase and the electronic anisotropy
order parameter $\eta$ introduced here. Symmetry arguments enforce
them to be proportional to each other, i.e. $\left\langle \delta\right\rangle \propto\left\langle \eta\right\rangle $.
However, the fact that $\delta$ is small in the HO phase ($\delta\sim10^{-5}$
as measured by x-ray diffraction \cite{Tonegawa14}) does not imply
that $\eta$ is also necessarily small. If indeed the anisotropy is
electronically driven, as suggested by the elasto-resistance measurements
\cite{IFisherNat15}, one would expect $\eta$ to be sizable even
if $\delta$ is small. This is the case, for instance, in optimally-doped
iron-based superconductors, where the in-plane resistivity anisotropy
is $\rho_{b}/\rho_{a}\sim1.5$ even though $\delta\sim10^{-4}$ (for
reviews, see \cite{Fisher_review_11,Fernandes14}).

\subsection{Nodal quasi-particle density of states}

Having established the existence of two SC states in $\mathrm{URu_{2}Si_{2}}$,
we now discuss their thermodynamic properties. At low temperatures,
they are determined by the low-energy properties of the quasi-particle
density of states (DOS) $\rho\left(\omega\right)$. For instance,
the specific heat $C$, the penetration depth $\lambda_{p}$, and
the spin-lattice relaxation rate $1/T_{1}T$ are given by:

\begin{align}
\frac{C}{T} & \propto\int d\omega\rho\left(\omega\right)\left(\beta\omega\right)^{2}\left(-\frac{\partial f}{\partial\omega}\right)\nonumber \\
\lambda_{p}^{-2}-\lambda_{p,0}^{-2} & \propto\int d\omega\rho\left(\omega\right)\left(-\frac{\partial f}{\partial\omega}\right)\nonumber \\
\left(T_{1}T\right)^{-1} & \propto\int d\omega\rho^{2}\left(\omega\right)\left(-\frac{\partial f}{\partial\omega}\right)\label{thermo}
\end{align}

At low energies, the DOS is determined by the dispersion of the nodal
quasi-particles. The latter are remarkably different for the two SC
states. For the higher-temperature SC state at $T_{c}^{*}<T<T_{c}$,
which we denote SC$_{1}$, only one of the gap components is condensed
and the gap function contains horizontal zeros at $k_{z}=0,\pm2\pi$
and vertical zeros at $k_{x}=-k_{y}$ (if $\Delta_{1}\neq0$) or at
$k_{x}=k_{y}$ (if $\Delta_{2}\neq0$). On the other hand, for the
lower-temperature SC state at $T<T_{c}^{*}$, which we denote SC$_{2}$,
the gap vanishes along the plane $k_{z}=0,\pm2\pi$ and along the
line $k_{x}=k_{y}=0$.

\begin{figure}[htbp]
\includegraphics[width=0.4\columnwidth]{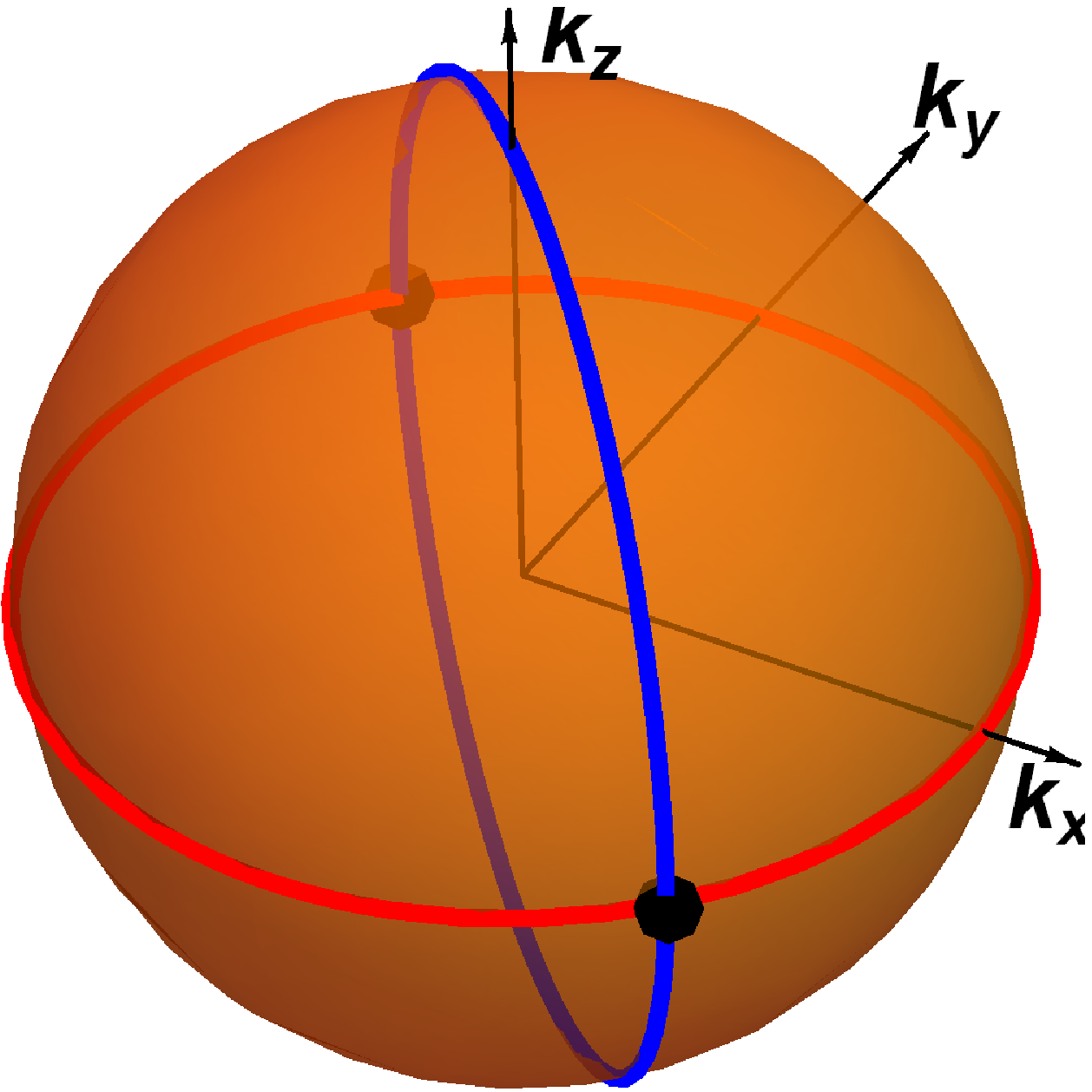}\hfill{} \includegraphics[width=0.4\columnwidth]{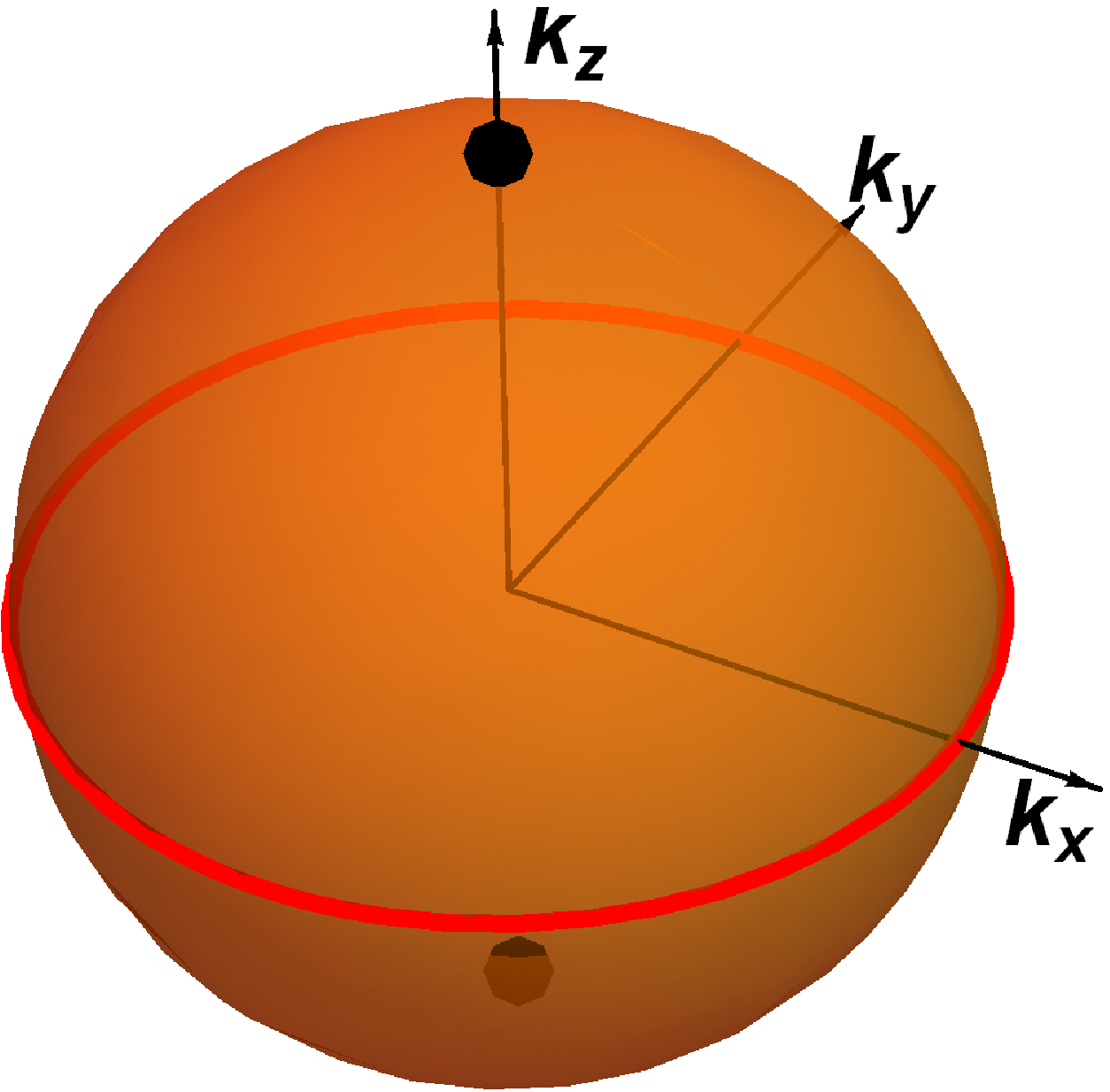}
\protect\caption{Nodes on a Fermi surface centered at $\Gamma$ in the two different
SC phases. Left: In the SC$_{1}$ phase, the Fermi surface contains
the horizontal nodal line $k_{z}=0$ (red curve) and the vertical
nodal line $k_{x}+k_{y}=0$ (blue curve). Right: in the SC$_{2}$
phase, the Fermi surface contains only one horizontal nodal line $k_{z}=0$
(red curve) and two nodal points $(0,0,\pm k_{F})$ (black dots). }

\label{Fig:SCNode}
\end{figure}

To proceed, we need to establish whether these gap zeros cross the
Fermi surface, forming nodal quasi-particles. The Fermi surface of
$\mathrm{URu_{2}Si_{2}}$ is remarkably complex: first-principle calculations
in the paramagnetic phase show the existence of hole-like ellipsoids
centered at $\Gamma=\left(0,0,0\right)$ and $Z=\left(0,0,2\pi\right)$,
and electron-like ellipsoids centered at $M=\left(\pi,\pi,0\right)$
\cite{Oppeneer10,Ikeda12}. Although it is unclear how the hybridization
with the $5f$ U states affects this Fermi surface configuration,
or whether there is an additional folding of the band structure along
the $\left(0,0,2\pi\right)$ momentum, as in the magnetically ordered
phase, quantum oscillation measurements \cite{Ohkuni99,Shishido09,Hassinger10,Altarawneh11,Tonegawa13}
seem to be consistent with at least one pocket centered at $\Gamma$.
For the purposes of determining the low-energy properties of the DOS,
we therefore consider a single spherical pocket centered at $\Gamma$.
In this case, in the SC$_{1}$ state, there are two crossing nodal
lines at $k_{F,z}=0$ and $k_{F,x}=\pm k_{F,y}$. In the SC$_{2}$
phase, there remains a nodal line at $k_{F,z}=0$ and a couple of
nodal points at $k_{F,x}=k_{F,y}=0$. Both nodal configurations are
shown in Fig.~\ref{Fig:SCNode}.

We first consider the higher-temperature SC$_{1}$ state ($T_{c}^{*}<T<T_{c}$).
For concreteness, we assume $\Delta_{1}\neq0$ and $\Delta_{2}=0$,
but the results are the same for the converse. The DOS is given by:
\begin{eqnarray}
\rho(\omega>0) & = & \frac{1}{V}\sum_{k}\delta(\omega-E(\vec{k}))\nonumber \\
 & = & N_{0}\int\frac{\rmd\Omega_{\hat{k}}}{4\pi}\int\rmd\xi\ \delta(\omega-\sqrt{\xi^{2}+\Delta^{2}(\hat{k})})\nonumber \\
 & = & N_{0}\omega\int\frac{\rmd\Omega_{\hat{k}}}{4\pi}\mathrm{Re}\frac{1}{\sqrt{\omega^{2}-\Delta^{2}(\hat{k})}}\ ,\label{def_DOS}
\end{eqnarray}
where $N_{0}$ is the DOS at the Fermi level, and $\Delta(\hat{k})$
is the SC gap along the Fermi pocket centered at $\Gamma$. The latter
can be conveniently described in terms of the polar and azimuthal
angles $\theta$ and $\phi$ around the spherical pocket:
\begin{eqnarray}
\Delta(\hat{k}) & = & \Delta_{1}\sin\frac{k_{z}}{2}\sin\frac{k_{x}+k_{y}}{2}\nonumber \\
 & \approx & \Delta_{1}\sin2\theta\cos\left(\phi-\frac{\pi}{4}\right)\ ,\label{gap_Delta_1}
\end{eqnarray}
where $\Delta_{1}$ has been rescaled by the factor $\Delta_{1}\rightarrow\Delta_{1}\sqrt{2}\left(k_{F}a\right)^{2}/8$.
There are two nodal lines at $\theta_{0}=\pi/2$ ($k_{z}=0$) and
$\phi_{0}=-\frac{\pi}{4},\,\frac{3\pi}{4}$ ($k_{x}=-k_{y}$ ), which
intersect along the Fermi surface at a single point, giving rise to
a quadratic node. Expanding the DOS for small energies $\omega\ll\Delta_{1}$,
we find the asymptotic behavior:
\begin{equation}
\rho(\omega)\approx N_{0}\left(\frac{\omega}{2\Delta_{1}}\log\frac{\Delta_{1}}{\omega}+\frac{\omega}{\Delta_{1}}\log4\right)\label{Eqn:DOSSC1}
\end{equation}

\begin{figure}[htbp]
\centering \subfigure[\label{Fig:DOSSC:SC1}]{\includegraphics[width=0.49\columnwidth]{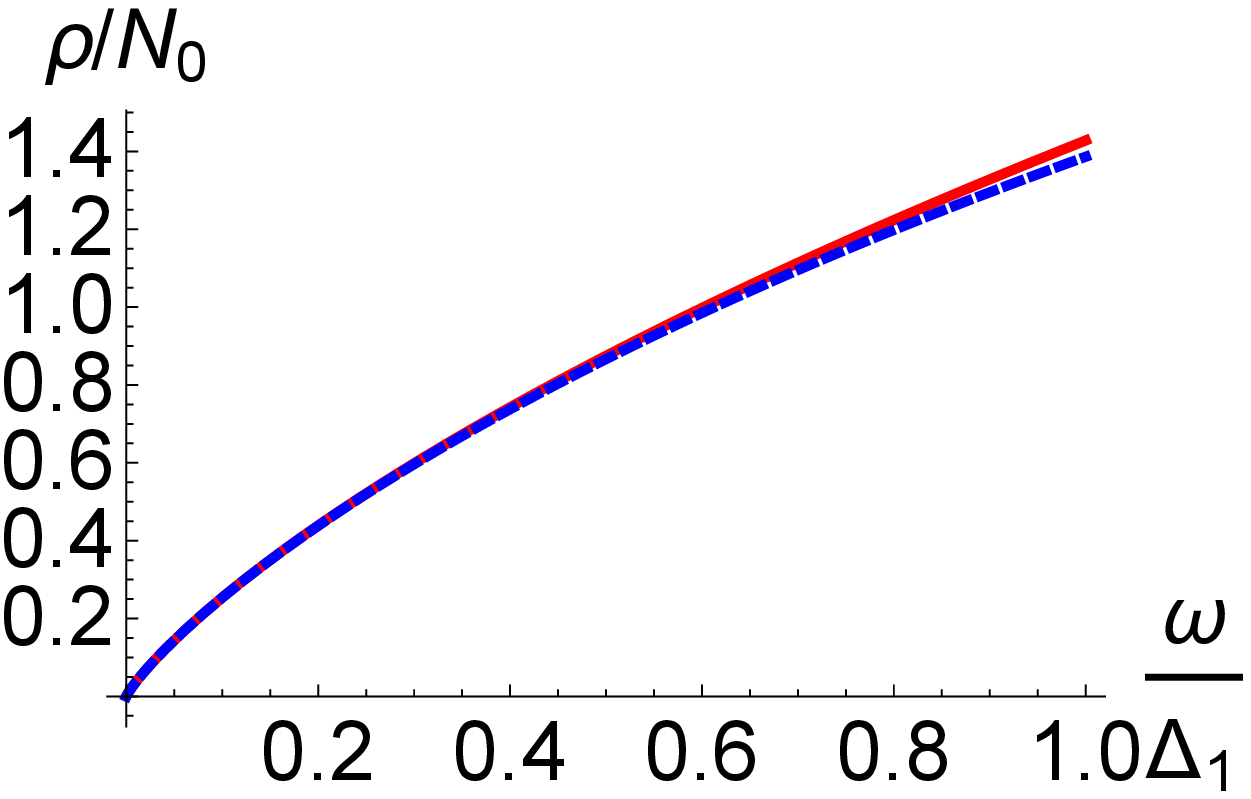}}
\subfigure[\label{Fig:DOSSC:SC2}]{\includegraphics[width=0.49\columnwidth]{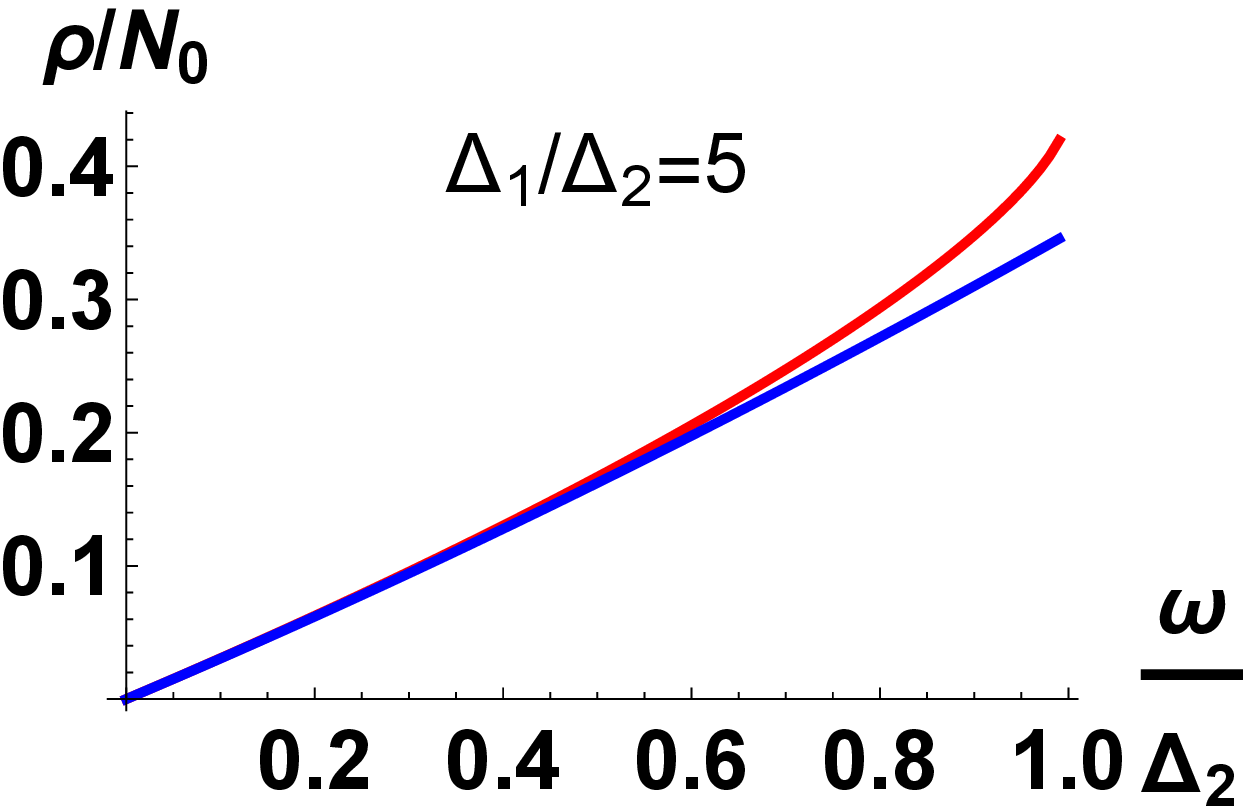}}
\protect\protect\protect\caption{The low energy DOS computed numerically (red curve) and via its asymptotic
expression (blue curve). Left: DOS in the high-temperature SC$_{1}$
phase ($T_{c}^{*}<T<T_{c}$). The blue curve is the asymptotic expression
\eqref{Eqn:DOSSC1} with the dominant $\omega\log\omega$ behavior
at low energies. Right: DOS in the low-temperature SC$_{2}$ phase
($T<T_{c}^{*}$). The blue curve is the asymptotic expression \eqref{Eqn:DOSSC2}
with the dominant $\omega$ behavior at low energies.}

\label{Fig:DOSSC}
\end{figure}

Fig.~\ref{Fig:DOSSC} compares the numerically-evaluated DOS with
the asymptotic expression (\ref{Eqn:DOSSC1}) as function of energy,
evidencing the dominant $\omega\log\omega$ behavior at low energies.

A careful analysis of Eq. (\ref{Eqn:DOSSC1}) reveals that while the
linear-in-$\omega$ term arises from the contribution of each individual
nodal line, the $\omega\log\omega$ term arises from the crossing
point between the two nodal lines. To make this transparent, we expand
the SC order parameter in the vicinity of one of the two crossing
points, $\theta_{0}=\pi/2$, $\phi_{0}=3\pi/4$:
\begin{equation}
\Delta\left(\theta,\phi\right)=2\Delta_{1}\left(\theta-\theta_{0}\right)\left(\phi-\phi_{0}\right)\ .
\end{equation}

Clearly, the nodal dispersion is quadratic rather than linear \cite{Fernandes11,Stanev11,Mazidian13,Kang14}.
Substituting this expression in Eq. (\ref{def_DOS}) and restricting
the integration to the proximities of the crossing point, $\left|\theta-\theta_{0}\right|,\left|\phi-\phi_{0}\right|<\Lambda$,
yields:

\begin{equation}
\rho(\omega)\approx4N_{0}\frac{\omega}{2\Delta_{1}}\int_{0}^{\Lambda}d\tilde{\theta}\int_{0}^{\Lambda}d\tilde{\phi}\,\mathrm{Re}\frac{1}{\sqrt{\left(\dfrac{\omega}{2\Delta_{1}}\right)^{2}-\tilde{\phi}^{2}\tilde{\theta}^{2}}}\ ,
\end{equation}
where $\tilde{\theta}=\theta-\theta_{0}$, $\tilde{\phi}=\phi-\phi_{0}$.
A straightforward calculation gives:

\begin{equation}
\rho(\omega)\approx\pi N_{0}\frac{\omega}{\Delta_{1}}\log\left(\frac{4\Lambda^{2}\Delta_{1}}{\omega}\right)\ .
\end{equation}

Clearly, the additional logarithmic contribution is a consequence
of the quadratic dispersion near the crossing point of two nodal lines.
Note that it will also give rise to logarithmic corrections to the
low-temperature behavior of the thermodynamic quantities listed in
the beginning of this section. More generally, tunneling experiments
sensitive to $\rho\left(\omega\right)$ may in principle be able to
identify this additional logarithmic contribution, which would be
unambiguous evidence for this type of superconducting gap function.

In the low-temperature SC$_{2}$ phase at $T<T_{c}^{*}$, $\Delta_{2}\neq0$
and one of the nodal lines is replaced by two nodal points. As a result,
the logarithmic corrections discussed above disappear. Indeed, the
system in the SC$_{2}$ phase contains only one nodal line $k_{z}=0$,
and two nodal points $k_{x}=k_{y}=0$. Computing the DOS at low energies,
$\omega\ll\Delta_{2},\Delta_{1}$, we find:
\begin{equation}
\rho(\omega)\approx N_{0}\left[\frac{\omega}{2\Delta_{1}}\ln\left(\frac{4\Delta_{1}}{\Delta_{2}}\right)+\frac{\omega^{2}}{4\Delta_{1}\Delta_{2}}\right]\label{Eqn:DOSSC2}
\end{equation}

As expected, the nodal line gives a linear contribution to the DOS,
whereas the nodal points give a quadratic contribution. Of course,
at energies higher than the scale of $\Delta_{2}$, but still much
smaller than $\Delta_{1}$, there is a crossover to the logarithmic
behavior found in the SC$_{1}$ phase, which may be detectable experimentally.

Interestingly, the coefficient of the linear-in-$\omega$ term depends
on the ratio $\Delta_{1}/\Delta_{2}$, which in turn depends on the
$C_{4}$ symmetry-breaking parameter $\eta$, according to Eq.~(\ref{SC2}).
The dependence of this linear coefficient on $\eta$ is shown in Fig.~\ref{Fig:LinDOSSC2}
below. This result reveals an interesting possibility to probe the
impact of the tetragonal symmetry breaking of the HO phase on the
SC properties. Specifically, one could experimentally extract this
linear coefficient -- either from tunneling experiments or from penetration
depth data -- and study its dependence on compressive and tensile
external strain, which would tend to increase or reduce the value
of $\eta$.

\begin{figure}[htbp]
\centering \includegraphics[width=0.7\columnwidth]{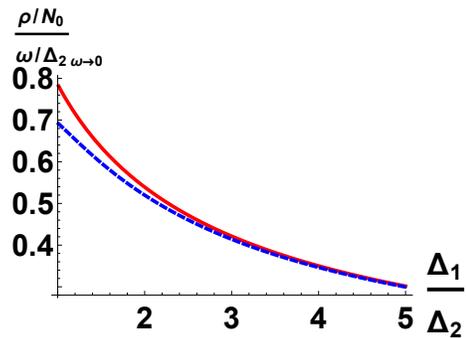} \protect\protect\caption{The linear coefficient of $\omega$ in the low energy DOS of the SC$_{2}$
phase as function of $\frac{\Delta_{1}}{\Delta_{2}}-1\propto\eta$.
The red curve is the numerical result and the blue curve corresponds
to the asymptotic expression \eqref{Eqn:DOSSC2}.}

\label{Fig:LinDOSSC2}
\end{figure}

\section{Anisotropic properties of the superconducting state}

Having established the thermodynamic properties of the SC$_{1}$ and
SC$_{2}$ phases, we now discuss how the in-plane anisotropy appearing
at the HO transition temperature is manifested in the superconducting
properties. From symmetry considerations, one expects anisotropies
in the in-plane penetration depth, critical magnetic field $H_{c2}$,
etc. Experimentally, the behavior of the specific heat as function
of the angle of an external magnetic field $\vec{H}$ has been used
as a probe of the nodal structure of the superconducting gap in URu$_{2}$Si$_{2}$
\cite{Yano08}. Thus, given the experimental feasibility and the existence
of current data, we focus here on the magnetic-field angle-dependent
specific heat, $C\big(\vec{H}\big)$, in both SC$_{1}$ and SC$_{2}$
phases. In the presence of an external magnetic field, a nodal superconductor
acquires a finite density of states $\rho\big(\vec{H}\big)$ at $T=0$.
As a result, the specific heat at low temperatures is given by $C\big(\vec{H}\big)/T\propto\rho\big(\vec{H}\big)$.
In a tetragonal system, $\rho\big(\vec{H}\big)$ displays four-fold
oscillations as the azimuthal angle of the applied magnetic field
$\vec{H}$ changes. However, in an orthorhombic system, which is the
case of URu$_{2}$Si$_{2}$ below the HO transition, two-fold oscillations
are expected.

A complete description of $\rho\big(\vec{H}\big)$, which is beyond
the scope of this work, requires detailed knowledge of the Fermi surface
and of the microscopic mechanisms of the superconducting state \cite{Vekhter01,Vekhter07}.
Here, we are interested in contrasting the anisotropies introduced
by $\eta$ and manifested in $\rho\big(\vec{H}\big)$ in both SC phases,
and also in studying the impact of twin domains present in real materials.
Therefore, to keep the discussion general and meaningful, we consider
a spherical Fermi surface and the semi-classical approximation for
$\rho\big(\vec{H}\big)$ first introduced by Volovik \cite{Volovik93}.
Valid in the regime of low fields and low temperatures, where the
vortex lattice of the mixed state is dilute, this approximation focuses
on the contributions of the extended quasi-particles around a single
vortex, neglecting the contribution from the vortex core states. The
main effect arises from the supercurrent $\vec{v}_{s}\left(\vec{r}\right)$
generated by an isolated vortex, which causes a Doppler shift $\Delta E=\vec{k}\cdot\vec{v}_{s}\left(\vec{r}\right)$
in the quasi-particle excitation spectrum. Taking the spatial average
over the unit cell of the vortex lattice then gives the zero-energy
DOS \cite{Volovik93,Vekhter01}:

\begin{equation}
\rho(\vec{H})=\frac{1}{V}\sum_{\vec{k}}\int\frac{\rmd^{2}r}{A}\delta\left(\vec{k}\cdot\vec{v}_{s}-E_{\vec{k}}\right)\label{Eqn:DOS_Doppler}
\end{equation}
where $A=\Phi_{0}/H=\pi R^{2}$ is the area of the vortex lattice
unit cell, which in turn is approximated by a circle of radius $R=\sqrt{\Phi_{0}/(\pi H)}$.
Here, $\Phi_{0}$ is the flux quantum and $E_{\vec{k}}=\sqrt{\xi_{\vec{k}}^{2}+\Delta^{2}\big(\vec{k}\big)}$
is the quasi-particle energy. For an arbitrary magnetic field along
the $\hat{n}$ direction, $\vec{H}=H\hat{n}$, the supercurrent velocity
$\vec{v}_{s}$ is given by:
\begin{equation}
\vec{v}_{s}=\frac{1}{2mr}\hat{n}\times\hat{r}\ .
\end{equation}

To proceed, it is convenient to define the angle-dependent Doppler-shift
energy:
\begin{equation}
E_{D}(\vec{k})=\frac{k_{f}}{2mR}\left|\hat{n}\times\hat{k}\right|\ ,
\end{equation}

Evaluation of the DOS in Eq. (\ref{Eqn:DOS_Doppler}) then gives:
\begin{equation}
\frac{\rho(\vec{H})}{N_{0}}=\int\frac{\rmd\Omega_{\hat{k}}}{4\pi}\left\{ \begin{array}{ll}
\dfrac{E_{D}^{2}(\hat{k})}{\Delta^{2}(\hat{k})} & \mbox{if }E_{D}(\hat{k})\leq|\Delta(\hat{k})|\\
1 & \mbox{if }E_{D}(\hat{k})\geq|\Delta(\hat{k})|
\end{array}\right.\label{eq_rho_0}
\end{equation}

It is convenient to define the ratio between the Doppler-shift energy
scale and the magnitude of the SC gap:

\begin{equation}
\gamma=\frac{k_{F}}{4mR\Delta_{1}}\ ,\label{eq_gamma}
\end{equation}

We consider first the SC$_{1}$ state, when $\Delta_{1}\neq0$ and
$\Delta_{2}=0$. Since we are interested mainly on the in-plane anisotropies
promoted by the HO tetragonal symmetry-breaking parameter $\eta$,
hereafter we consider the situation in which $\vec{H}$ is swept across
the $ab$ plane, and therefore characterized by the azimuthal angle
$\phi_{h}$. Using the expansion (\ref{gap_Delta_1}) for $\Delta_{1}$
yields the expression:

\begin{equation}
\dfrac{E_{D}(\hat{k})}{2\left|\Delta(\hat{k})\right|}=\gamma\,\frac{\sqrt{1-\sin^{2}\theta\cos^{2}\left(\phi-\phi_{h}\right)}}{\left|\sin2\theta\cos\left(\phi-\frac{\pi}{4}\right)\right|}
\end{equation}
which can be substituted in Eq. (\ref{eq_rho_0}) for numerical evaluation.
In the weak field limit, $\gamma\ll1$, we obtain the analytical expression:

\begin{equation}
\frac{\rho(\phi_{h})}{4N_{0}}\approx\gamma+\left|\cos\left(\phi_{h}-\frac{\pi}{4}\right)\right|\gamma\left(\frac{1}{\pi}\ln\frac{2}{\gamma}-1\right)
\end{equation}

\begin{figure}[htbp]
\centering \subfigure[\label{Fig:DOSSC1:D01Hab}]{\includegraphics[width=0.6\columnwidth]{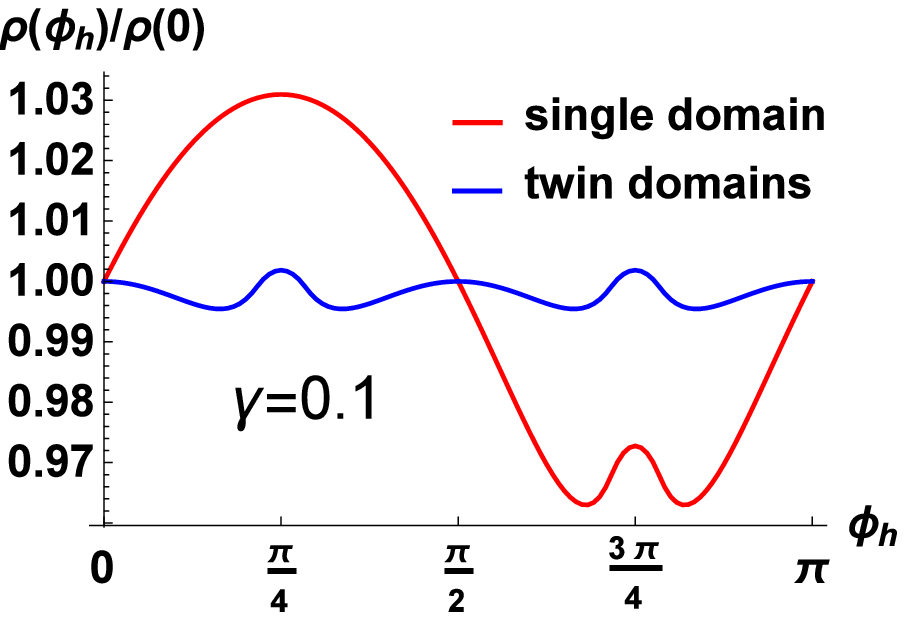}}
\subfigure[\label{Fig:DOSSC1:D001Hab}]{\includegraphics[width=0.6\columnwidth]{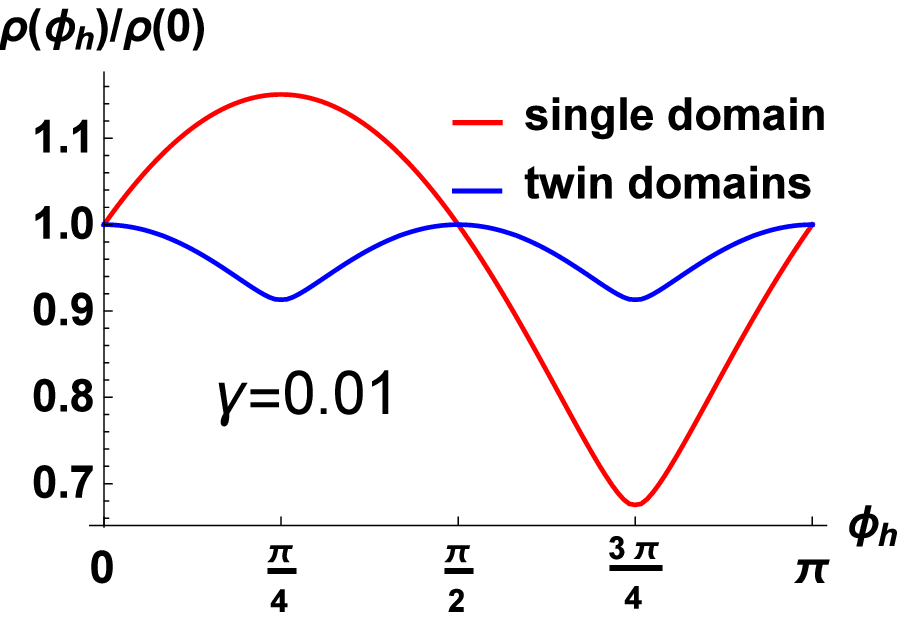}}
\protect\protect\protect\protect\caption{The angle-dependent DOS in the presence of an in-plane magnetic field
in the SC$_{1}$ state. The ratio between the Doppler shift energy
and the gap is set to $\gamma=0.1$ in panel (a) and $\gamma=0.01$
in panel (b). $\phi_{h}$ is the azimuthal angle of the magnetic field.
The red curve corresponds to a single domain with $\Delta_{2}=0$,
whereas the blue curve corresponds to a twin domain with equal-weight
$\Delta_{2}\protect\neq0,\,\Delta_{1}=0$ and $\Delta_{1}\protect\neq0,\,\Delta_{2}=0$. }

\label{Fig:DOSSC1_H}
\end{figure}

This formula captures the general behavior of the numerically-calculated
$\rho\left(\phi_{h}\right)$, plotted in Fig.~\ref{Fig:DOSSC1_H}.
In particular, it reveals a clear in-plane two-fold anisotropy of
$\rho(\phi_{h})$, with a maximum at $\phi_{h}=\pi/4,\,-3\pi/4$ and
a minimum at $\phi_{h}=-\pi/4,\,3\pi/4$. This behavior is a consequence
of the quadratic node present in the Fermi surface at $\phi_{0}=\pi/4$,
$\theta_{0}=\pi/2$ (see the previous Section): when $\vec{H}$ is
parallel to the quadratic node momentum, the Doppler shift in the
nodal quasi-particle spectrum is maximum, but when $\vec{H}$ is perpendicular
to the quadratic node momentum, the Doppler shift is minimum.

\begin{figure}[htbp]
\centering \subfigure[\label{Fig:DOSSC2:D01Hab}]{\includegraphics[width=0.6\columnwidth]{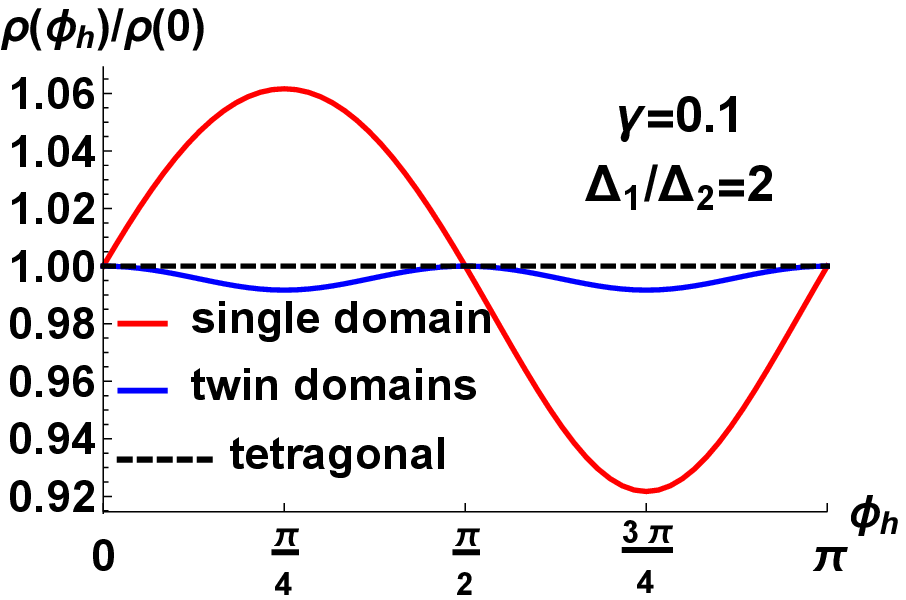}}
\subfigure[\label{Fig:DOSSC2:D001Hab}]{\includegraphics[width=0.6\columnwidth]{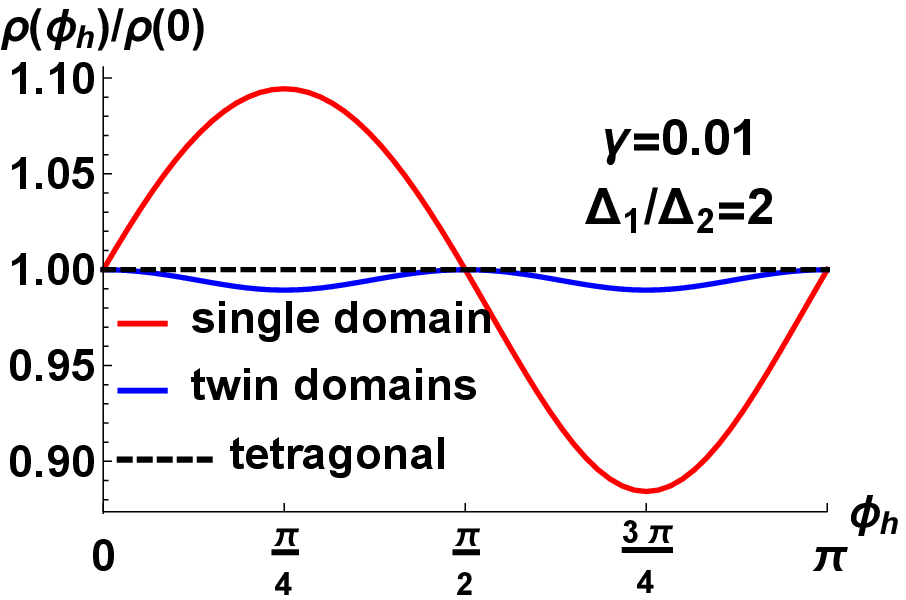}}
\protect\protect\caption{The angle-dependent DOS in the presence of an in-plane magnetic field
in the SC$_{2}$ state ($\Delta_{1},\Delta_{2}\protect\neq0$). The
ratio between the Doppler shift energy and the gap is set to $\gamma=0.1$
in the upper panel and $\gamma=0.01$ in the lower panel. $\phi_{h}$
is the azimuthal angle of the magnetic field. The red curve corresponds
to a single domain with $\Delta_{1}/\Delta_{2}=2$, whereas the blue
curve corresponds to a twin domain with equal-weight $\Delta_{1}/\Delta_{2}=2$
and $\Delta_{1}/\Delta_{2}=1/2$. The dashed black curve corresponds
to the tetragonal symmetric system, with $\Delta_{1}=\Delta_{2}$.}

\label{Fig:DOSSC2_H}
\end{figure}

We now consider the SC$_{2}$ phase, which is the most relevant for
comparison with experiments, since the latter are commonly performed
at very low temperatures -- presumably below both $T_{c}$ and $T_{c}^{*}$.
In the SC$_{2}$ phase, both $\Delta_{1}$ and $\Delta_{2}$ are non-zero,
but non-equal. We obtain:

\begin{equation}
\dfrac{E_{D}(\hat{k})}{2\left|\Delta(\hat{k})\right|}=\gamma\,\frac{\sqrt{1-\sin^{2}\theta\cos^{2}\left(\phi-\phi_{h}\right)}}{\left|\sin2\theta\right|\sqrt{\cos^{2}\left(\phi-\frac{\pi}{4}\right)+\frac{\Delta_{2}^{2}}{\Delta_{1}^{2}}\sin^{2}\left(\phi-\frac{\pi}{4}\right)}}
\end{equation}
which can then be substituted in Eq. (\ref{eq_rho_0}) to compute
the DOS. In Fig. \ref{Fig:DOSSC2_H}, we plot $\rho\left(\phi_{h}\right)$
for two different values of the parameter $\gamma$, considering the
ratio $\Delta_{1}/\Delta_{2}=2$. The behavior is similar to the SC$_{1}$
phase, namely, the DOS is two-fold anisotropic and maximum at $\phi_{h}=\pi/4,\,-3\pi/4$
but minimum at $\phi_{h}=-\pi/4,\,3\pi/4$. The contrast between the
maximum and the minimum is proportional to the ratio $\Delta_{1}/\Delta_{2}$.
In the $C_{4}$ symmetric case, $\Delta_{1}=\Delta_{2}$, to leading
order in $\gamma$, $\rho\left(\phi_{h}\right)$ is angle-independent
and constant, $\rho\left(\phi_{h}\right)=4\gamma/\pi+\gamma^{2}\ln(2/\gamma)$.

An important issue is whether the two-fold anisotropy induced by the
difference between $\Delta_{1}$ and $\Delta_{2}$ (which in turn
arises from the HO $C_{4}$ symmetry-breaking term $\eta$) can be
observed experimentally. In our analysis, so far we have considered
only the ideal scenario in which a single $C_{2}$ domain is formed.
In large samples -- at least large enough to allow one to measure
the specific heat -- it is very likely that the system will break
up in twin domains with $\eta>0$ (corresponding to $\Delta_{2}=0$
in the SC$_{1}$ phase and $\Delta_{1}>\Delta_{2}$ in the SC$_{2}$
phase) and $\eta<0$ (corresponding to $\Delta_{1}=0$ in the SC$_{1}$
phase and $\Delta_{1}<\Delta_{2}$ in the SC$_{2}$ phase). As a result,
the two-fold anisotropy is washed out. To capture this effect, we
calculated $\rho_{t}\left(\phi_{h}\right)$ of a twin-domain system:

\begin{equation}
\rho_{t}(\phi_{h})=\frac{1}{2}\rho(\phi_{h})+\frac{1}{2}\rho\left(\phi_{h}+\pi/2\right)\ .
\end{equation}

The results for both SC$_{1}$ and SC$_{2}$ phases in the twin-domain
case are shown by the blue curves in Figs. \ref{Fig:DOSSC1_H} and
\ref{Fig:DOSSC2_H}. While for the SC$_{1}$ phase there is a small
four-fold anisotropy reminiscent of the anisotropies of the single-domain
case, in the SC$_{2}$ phase the DOS is basically angle-independent
and indistinguishable from the tetragonal case ($\Delta_{1}=\Delta_{2}$).
Interestingly, the in-plane angle-resolved specific heat data reported
in \cite{Yano08} is nearly flat as the field is swept from the $a$
axis to the $b$ axis.

\begin{figure}[htbp]
\centering \includegraphics[width=0.6\columnwidth]{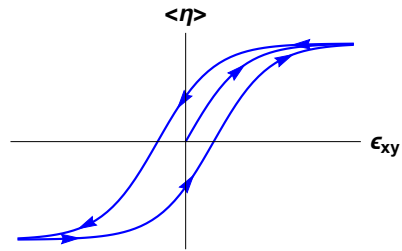} \protect\protect\protect\caption{Bulk averaged anisotropy $\langle\eta\rangle$ as function of the
in-plain uniaxial strain $\varepsilon_{xy}$. Without uniaxial strain,
$\langle\eta\rangle=0$ as averaged over different domains. It becomes
nonzero upon application of uniaxial strain and remains non-zero even
after the strain is released, inside the symmetry-broken phase.}

\label{Fig:AniStrain}
\end{figure}

In order to decide whether the experimentally observed behavior is
compatible with the orthorhombic SC$_{2}$ phase or with the tetragonal
SC chiral phase, we propose to perform angle-resolved specific heat
experiments in samples under uniaxial external strain. In both cases,
applying either compressive or tensile stress along the $a$ (or $b$)
axis will induce a two-fold anisotropy in $\rho\left(\phi_{h}\right)$
-- similarly to what we calculated for a single-domain case. This
is not surprising, as the external strain explicitly breaks the $C_{4}$
symmetry -- in other words, it generates itself a term $\eta$ in
the free energy (\ref{Eqn:Free}) regardless of whether the symmetry
was spontaneously broken at the HO transition. However, upon releasing
the strain, only in the SC$_{2}$ phase a non-zero $\eta$ remains
due to the alignment of the twin domains: This is nothing but the
manifestation of the hysteresis associated with the symmetry-broken
phase (see Fig.~\ref{Fig:AniStrain}). Therefore, if the two-fold
anisotropies of $\rho\left(\phi_{h}\right)$ persist after the applied
strain is released, this would be unambiguous proof of the anisotropy
in $\Delta_{1}$ and $\Delta_{2}$ characteristic of the SC$_{2}$
phase. Note that devices capable of applying and releasing strain
in a controlled way have been now widely used to study the nematic
phase of the iron-based superconductors \cite{Degiorgi15}.

\section{Collective modes in the superconducting phase}

As discussed above, inside the HO phase there are two SC transitions:
at $T_{c}$, the system undergoes a normal-state to SC transition,
characterized by the condensation of one of the two components of
the chiral SC order parameter ($\Delta_{1}$ if $\eta>0$). At $T_{c}^{*}<T_{c}$,
a SC-to-SC transition takes place, in which the second component of
the SC order parameter condenses ($\Delta_{2}$ if $\eta>0$) with
a $\pi/2$ relative phase with respect to the other component, breaking
time-reversal symmetry. Detection of the second SC transition in URu$_{2}$Si$_{2}$
would be strong evidence for the scenario discussed here. Besides
the usual thermodynamic signatures of a SC transition, for instance
in the specific heat, the system also displays distinctive collective
SC excitations at $T_{c}^{*}$ -- which in turn could be measured
by spectroscopic techniques such as Raman scattering.

To understand why this is the case, we note that the collective modes
of a single-band superconductor are the phase and amplitude (also
called Higgs) modes. While the former is always gapped by the Anderson-Higgs
mechanism, the latter becomes soft near the superconducting transition.
Near the usual normal-state to SC transition (such as the one taking
place at $T_{c}$ in our system), the energy gap of the amplitude
mode is generally larger than the SC gap, implying that the mode decays
into the particle-hole continuum. However, near a SC-to-SC transition
(such as the one taking place at $T_{c}^{*}$ in our system) the situation
is different, because the electronic spectrum has already been gapped
by the first SC transition. As a result, it becomes in principle possible
to obtain a sharp SC amplitude mode at $T_{c}^{*}$ that does not
fall into the particle-hole continuum \cite{Larkin, Babaev,Stanev12,Maiti13}.
The situation is similar to the appearance of the so-called Bardasis-Schrieffer
mode in superconductors with nearly-degenerate superconducting states
\cite{Bardasis}.

To investigate this scenario, we include the time-dependence of the
SC order parameter in the free energy (\ref{Eqn:Free}):
\begin{equation}
\tilde{F}_{\mathrm{SC}}\left(\Delta_{i},\frac{\partial\Delta_{i}}{\partial t}\right)=\sum_{i=1}^{2}\gamma_{\mathrm{sc}}\left|\frac{\partial\Delta_{i}}{\partial t}\right|^{2}-F_{\mathrm{SC}}\left(\Delta_{i}\right)\ ,\label{Eqn:Action}
\end{equation}
where $F_{\mathrm{SC}}$ is given by Eq.~\eqref{Eqn:Free} and the
coefficient of the time-dependent term is $\gamma_{\mathrm{sc}}>0$.
Because we are interested in the behavior below $T_{c}$, where the
SC system is particle-hole symmetric, the time-dependent term must
be quadratic in the time-derivative, in contrast to the usual linear
$i\Delta^{*}\frac{\partial\Delta}{\partial t}$ term that appears
in the Gross-Pitaevskii equation, where particle-hole symmetry is
not necessarily present\cite{Larkin,Varma15}.

To calculate the collective modes of the system, we linearize the
gap equations near $T_{c}$ by writing $\Delta_{i}=\Delta_{i,0}+\delta\Delta_{i}$,
where $\Delta_{i,0}$ are the solutions of the time-independent equations,
i.e. $\left(\frac{\partial F_{\mathrm{SC}}}{\partial\Delta_{i}}\right)_{\Delta_{i,0}}=0$.
We obtain the coupled linear equations:

\begin{equation}
-\gamma_{\mathrm{sc}}\frac{\partial^{2}}{\partial t^{2}}\left(\begin{array}{c}
\delta\Delta_{1}\\
\delta\Delta_{1}^{*}\\
\delta\Delta_{2}\\
\delta\Delta_{2}^{*}
\end{array}\right)=\boldsymbol{\Lambda}\left(\Delta_{i,0}\right)\left(\begin{array}{c}
\delta\Delta_{1}\\
\delta\Delta_{1}^{*}\\
\delta\Delta_{2}\\
\delta\Delta_{2}^{*}
\end{array}\right)\label{TDGL}
\end{equation}
with the matrix:
\begin{widetext}
\begin{equation}
\boldsymbol{\Lambda}\left(\Delta_{i}\right)=\begin{pmatrix}\frac{a-\eta}{2}+u|\Delta_{1}|^{2}+\frac{\beta}{2}|\Delta_{2}|^{2} & \frac{u}{2}\Delta_{1}^{2}+\frac{\alpha}{2}\Delta_{2}^{2} & \frac{\beta}{2}\Delta_{1}\Delta_{2}^{*}+\alpha\Delta_{1}^{*}\Delta_{2} & \frac{\beta}{2}\Delta_{1}\Delta_{2}\\
\frac{u}{2}\Delta_{1}^{*^{2}}+\frac{\alpha}{2}\Delta_{2}^{*^{2}} & \frac{a-\eta}{2}+u|\Delta_{1}|^{2}+\frac{\beta}{2}|\Delta_{2}|^{2} & \frac{\beta}{2}\Delta_{1}^{*}\Delta_{2}^{*} & \frac{\beta}{2}\Delta_{1}^{*}\Delta_{2}+\alpha\Delta_{1}\Delta_{2}^{*}\\
\frac{\beta}{2}\Delta_{1}^{*}\Delta_{2}+\alpha\Delta_{1}\Delta_{2}^{*} & \frac{\beta}{2}\Delta_{1}\Delta_{2} & \frac{a+\eta}{2}+u|\Delta_{2}|^{2}+\frac{\beta}{2}|\Delta_{1}|^{2} & \frac{u}{2}\Delta_{2}^{2}+\frac{\alpha}{2}\Delta_{1}^{2}\\
\frac{\beta}{2}\Delta_{1}^{*}\Delta_{2}^{*} & \frac{\beta}{2}\Delta_{1}\Delta_{2}^{*}+\alpha\Delta_{1}^{*}\Delta_{2} & \frac{u}{2}\Delta_{2}^{*^{2}}+\frac{\alpha}{2}\Delta_{1}^{*^{2}} & \frac{a+\eta}{2}+u|\Delta_{2}|^{2}+\frac{\beta}{2}|\Delta_{1}|^{2}
\end{pmatrix}
\end{equation}

\end{widetext}

Because we are interested only in the energy gap of the collective
modes, and not in their dispersions, we ignore the spatial dependence
of the SC gap. The energies of the four collective modes are therefore
given by:

\begin{equation}
\omega_{i}=\sqrt{\frac{\lambda_{i}}{\gamma_{\mathrm{sc}}}}
\end{equation}
where $\lambda_{i}$ are the eigenvalues of the matrix $\boldsymbol{\Lambda}$.
We first focus in the SC$_{1}$ phase, $T_{c}^{*}<T<T_{c}$, in which
the equilibrium gap functions $\Delta_{i,0}$ are given by Eqs.~(\ref{SC1}).
Diagonalizing the matrix $\boldsymbol{\Lambda}$ gives the following
eigenvalues:

\begin{align}
\lambda_{1} & =0\nonumber \\
\lambda_{2} & =-a+\eta\nonumber \\
\lambda_{3} & =\frac{a\left(u+\alpha-\beta\right)+\eta\left(u-\alpha+\beta\right)}{2u}\nonumber \\
\lambda_{4} & =\frac{a\left(u-\alpha-\beta\right)+\eta\left(u+\alpha+\beta\right)}{2u}\label{egvlue1}
\end{align}
with the corresponding eigenvectors:

\begin{align}
\mathbf{v}_{1}^{T} & =\left(\begin{array}{cccc}
-1 & 1 & 0 & 0\end{array}\right)\nonumber \\
\mathbf{v}_{2}^{T} & =\left(\begin{array}{cccc}
1 & 1 & 0 & 0\end{array}\right)\nonumber \\
\mathbf{v}_{3}^{T} & =\left(\begin{array}{cccc}
0 & 0 & 1 & -1\end{array}\right)\nonumber \\
\mathbf{v}_{4}^{T} & =\left(\begin{array}{cccc}
0 & 0 & 1 & 1\end{array}\right)\label{egvctor1}
\end{align}

Clearly, the two SC components $\Delta_{1}$ and $\Delta_{2}$ decouple
in this regime. The expressions for the eigenvectors reveal that the
eigenvalue $\lambda_{1}$ corresponds to the phase mode of the condensed
SC component. Although in our calculation it vanishes, it becomes
a massive mode once the coupling to the electronic density is included
(Anderson-Higgs mechanism) \cite{Anderson58,Varma82,Cote93}. The
eigenvalue $\lambda_{2}$ is the amplitude mode of the $\Delta_{1}$
gap. Although it becomes soft at $T_{c}$ , since $a_{c}=\eta$, it
does not give rise to a sharp collective mode because at $T_{c}$
the electronic spectrum is barely gapped, implying that the mode falls
into the particle-hole continuum.

The other two eigenvalues correspond to modes of the incipient $\Delta_{2}$
component, which condenses only at $T_{c}^{*}$. Because it condenses
with a relative phase of $\pi/2$ with respect to $\Delta_{1}$, $\lambda_{3}$
corresponds to an incipient amplitude mode, whereas $\lambda_{4}$
corresponds to an incipient relative phase mode. Because the latter
does not couple directly to the electronic density, it does not become
massive below $T_{c}^{*}$ \cite{Leggett66}. Interestingly, we observe
that at $T_{c}^{*}$, which corresponds to:

\begin{equation}
a_{*}=-\eta\left(\frac{u+\beta-\alpha}{u-\beta+\alpha}\right)\ ,
\end{equation}
the mode $\lambda_{3}$ becomes soft. In this case, because $\Delta_{1}$
has already been condensed at $T_{c}$, the electronic spectrum is
gapped, implying that this collective excitation can be a sharp mode
in the vicinity of $T_{c}^{*}$. To analyze how these modes evolve
below $T_{c}^{*}$, we diagonalize the matrix $\boldsymbol{\Lambda}$
in the SC$_{2}$ phase, $T\leq T_{c}^{*}$, in which $\Delta_{i,0}$
is given by Eq. (\ref{SC2}). We obtain:

\begin{align}
\lambda_{1} & =0\nonumber \\
\lambda_{2} & =\dfrac{-au}{u+\beta-\alpha}+\sqrt{\eta^{2}\left(\dfrac{u+\beta-\alpha}{u-\beta+\alpha}\right)+\dfrac{a^{2}(\beta-\alpha)^{2}}{(u+\beta-\alpha)^{2}}}\nonumber \\
\lambda_{3} & =\dfrac{-au}{u+\beta-\alpha}-\sqrt{\eta^{2}\left(\dfrac{u+\beta-\alpha}{u-\beta+\alpha}\right)+\dfrac{a^{2}(\beta-\alpha)^{2}}{(u+\beta-\alpha)^{2}}}\nonumber \\
\lambda_{4} & =\dfrac{-2a\alpha}{u+\beta-\alpha}\label{egvlue2}
\end{align}

\begin{figure}[htbp]
\centering \includegraphics[width=0.7\columnwidth]{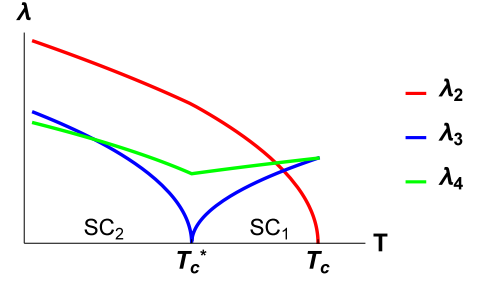} \protect\protect\protect\caption{Eigenmodes of the SC state inside the HO phase, as explained in the
main text: $\lambda_{2}$ (red), $\lambda_{3}$ (blue), and $\lambda_{4}$
(green). $\lambda_{1}$, corresponding to the global phase mode, is
not shown, since it becomes massive due to the coupling to the electronic
density.}

\label{Fig:Mode}
\end{figure}

Although the corresponding eigenvectors are straightforward to obtain,
we refrain from writing explicitly their lengthy expressions here.
In Fig.~\ref{Fig:Mode}, we plot the eigenvalues $\lambda_{2}$,
$\lambda_{3}$, and $\lambda_{4}$ as function of temperature inside
the SC state. As discussed above, $\lambda_{3}$, corresponding to
the incipient $\Delta_{2}$ amplitude mode, vanishes at $T_{c}^{*}$.
This opens the interesting possibility of detecting the second SC
transition spectroscopically. For instance, Raman scattering in the
symmetry channel corresponding to the $\Delta_{2}$ component ($B_{3g}$/$B_{2g}$
irreducible representation of the orthorhombic point group) could
in principle detect a sharp mode near $T_{c}^{*}$ inside the gapped
region of the spectrum ($\omega<2\Delta_{1}$).

\begin{figure}[htbp]
\centering \subfigure[\label{Fig:SelfEne:Diagram}]{\includegraphics[width=0.55\columnwidth]{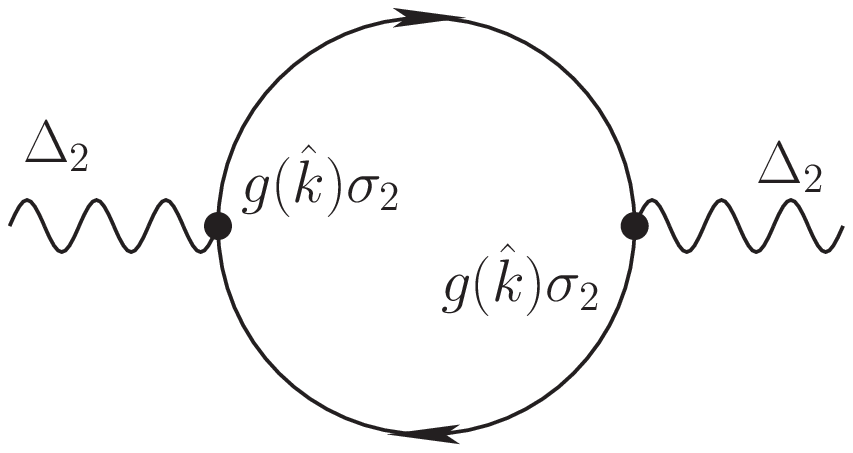}}
\subfigure[\label{Fig:SelfEne:Pi}]{\includegraphics[width=0.6\columnwidth]{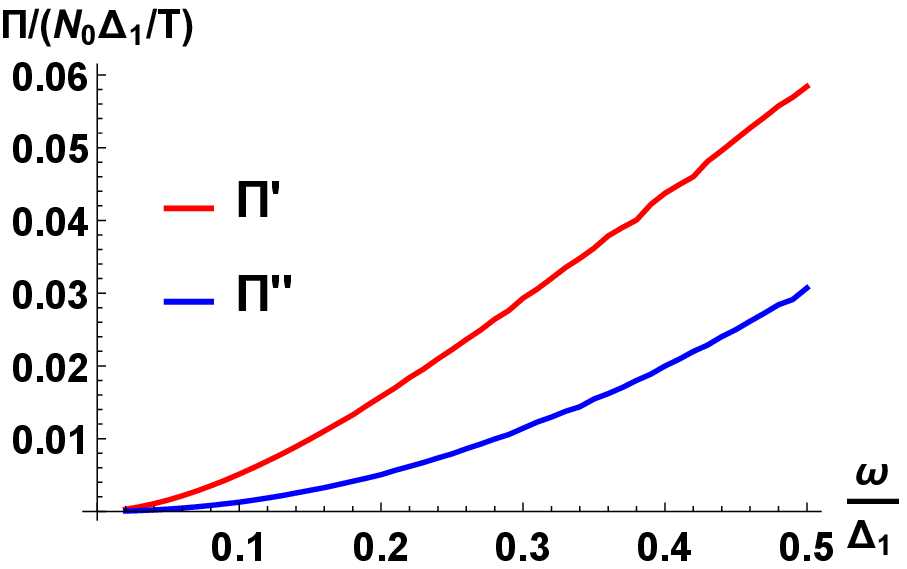}}
\subfigure[\label{Fig:SelfEne:PiRatio}]{\includegraphics[width=0.6\columnwidth]{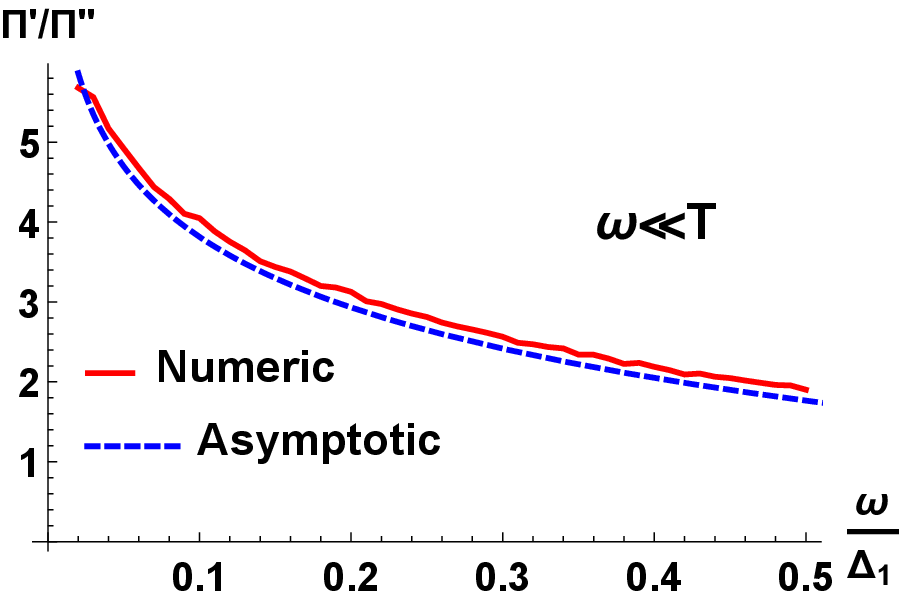}}
\protect\protect\protect\caption{Upper panel: Bosonic SC self-energy (in Nambu space) arising from
the coupling between $\Delta_{2}$ and the quasi-particles near $T_{c}^{*}$.
Here $g(\hat{k})=\sin(2\theta)\sin(\phi-\pi/4)$. Middle panel: the
real and imaginary parts of the bosonic self energy when $\omega\ll T$.
Lower panel: The ratio between the real and the imaginary parts of
the frequency-dependent bosonic self-energy. It is clear that the
real part is much smaller than the imaginary part at low frequencies.
Moreover, since the imaginary part is proportional to $\omega^{2}$
and not to $\omega$, the mode is under-damped.}

\label{Fig:SelfEne}
\end{figure}

An important issue ignored in the analysis above is the fact that
the superconducting state has nodal quasi-particle excitations, which
in principle couple to the $\lambda_{3}$ mode and can cause damping.
To investigate this effect, we compute the one-loop bosonic self-energy
diagram containing the coupling of $\Delta_{2}$ to the electronic
states (and particularly to the nodal quasi-particles) at the vicinity
of $T_{c}^{*}$, where the $\lambda_{3}$ mode becomes soft (see Fig.
\ref{Fig:SelfEne}). At this temperature, the SC gap $\Delta_{1}$
is fully developed. Therefore, in Nambu space, the electronic Green's
function is given by:

\begin{equation}
G(\vec{k},i\nu_{n})=\int\frac{\rmd z}{2\pi}\frac{A(\vec{k},z)}{i\nu_{n}-z}\ ,
\end{equation}
with $\nu_{n}=\left(2n+1\right)\pi T$ and the spectral function:
\begin{align}
A(\vec{k},z) & =\frac{\pi}{E_{\vec{k}}}\big(\delta(z-E_{\vec{k}})-\delta(z+E_{\vec{k}})\big)\nonumber \\
 & \times(z\sigma_{0}+\xi_{\vec{k}}\sigma_{3}-\Delta_{1}(\vec{k})\sigma_{1})
\end{align}

where $E_{\vec{k}}=\sqrt{\xi_{\vec{k}}^{2}+\Delta_{1}^{2}(\vec{k})}$
is the nodal quasi-particle excitation and $\sigma_{i}$ are Pauli
matrices in Nambu space. Thus, the bosonic self-energy shown in Fig.~\ref{Fig:SelfEne}
becomes:

\begin{align}
\Pi\left(i\omega_{n}\right) & =-T\sum_{m}\int\frac{\rmd^{d}k}{\left(2\pi\right)^{d}}\int\frac{dz_{1}}{2\pi}\int\frac{dz_{2}}{2\pi}\nonumber \\
 & \times\frac{\pi^{2}}{E_{\vec{k}}^{2}}\frac{\mathrm{tr}\left[g(\hat{k})\sigma_{2}A\left(\vec{k},z_{1}\right)g(\hat{k})\sigma_{2}A\left(\vec{k},z_{2}\right)\right]}{\left(i\nu_{m}+i\omega_{n}-z_{1}\right)\left(i\nu_{m}-z_{2}\right)}
\end{align}

Here we set the external momentum $q=0$ (since we are only interested
in the dynamics) and included the Nambu vertex $g(\hat{k})\sigma_{2}=\sin(2\theta)\sin(\phi-\pi/4)\sigma_{2}$
corresponding to $\Delta_{2}(\vec{k})=\Delta_{2}g(\hat{k})\sigma_{2}$.
A straightforward calculation gives:

\begin{align}
\Pi(i\omega_{n}) & =\int\frac{\rmd^{d}k}{(2\pi)^{d}}\tanh\left(\frac{\beta E_{\vec{k}}}{2}\right)\nonumber \\
 & \left(\frac{1}{i\omega+2E_{\vec{k}}}-\frac{1}{i\omega-2E_{\vec{k}}}\right)g^{2}(\hat{k})
\end{align}

Performing the analytic continuation and subtracting the frequency-independent
part, $\Pi\left(0\right)$, we obtain the imaginary and real parts
of $\delta\Pi\left(\omega\right)=\Pi\left(\omega\right)-\Pi\left(0\right)$
(hereafter we consider $\omega>0$):

\begin{align}
\delta\Pi''(\omega) & =\pi\tanh\left(\frac{\beta\omega}{4}\right)\int\frac{\rmd^{d}k}{(2\pi)^{d}}\delta(\omega-2E_{\vec{k}})g^{2}(\hat{k})\nonumber \\
\delta\Pi'(\omega) & =\int\frac{\rmd^{d}k}{(2\pi)^{d}}\tanh\left(\frac{\beta E_{\vec{k}}}{2}\right)\frac{\omega^{2}g^{2}(\hat{k})}{E_{\vec{k}}\left(4E_{\vec{k}}^{2}-\omega^{2}\right)}\label{eq_Pi}
\end{align}

In the limit $\omega\ll\Delta_{1},T$, we find the low-energy asymptotic
behaviors of the bosonic self-energy:
\begin{align}
\delta\Pi''(\omega) & \approx\frac{\pi N_{0}}{24}\frac{\omega^{2}}{T\Delta_{1}}\nonumber \\
\delta\Pi'(\omega) & \approx\frac{N_{0}}{6}\frac{\omega^{2}}{T\Delta_{1}}\log\left(\frac{2\Delta_{1}}{\omega}\right)
\end{align}

Thus, at low enough frequencies, not only does the imaginary part
varies quadratically with the frequency, but also the real part is
much larger than the imaginary part. Consequently, the $\lambda_{3}$
mode can still be sharp near $T_{c}^{*}$ despite the damping introduced
by its coupling to the nodal quasi-particles. To confirm these analytical
results, in Fig. \ref{Fig:SelfEne}, we plot the behavior of $\delta\Pi''\left(\omega\right)$
and $\delta\Pi'\left(\omega\right)$ evaluated numerically from Eqs.
(\ref{eq_Pi}), evidencing the sub-leading character of the imaginary
part.

\section{Concluding remarks}

In summary, we have investigated the impact of the tetragonal symmetry
breaking promoted by the HO phase in the low-temperature chiral SC
state of URu$_{2}$Si$_{2}$. Besides the anticipated splitting of
the SC transition into two, the two resulting SC phases display very
different low-energy behaviors. In particular, the nodal quasi-particle
density of states of the higher-temperature SC phase acquires an anomalous
logarithmic dependence due to the crossing of two nodal lines. Although
absent in the lower-temperature SC phase at low energies, this log-behavior
can in principle still be manifested for intermediate energy ranges
as a crossover effect. We have also shown the softening of one of
the amplitude SC modes near the second SC transition, providing yet
another signature of the interplay between tetragonal symmetry breaking
and SC. Finally, we showed that the current angle-resolved specific-heat
data is qualitatively consistent with either a tetragonal chiral state
or an orthorhombic chiral state. We propose additional measurements
in the presence of uniaxial strain to unambiguously distinguish the
two scenarios. 

It is important to critically analyze our results in face of recent
data on the SC state of URu$_{2}$Si$_{2}$. First, our phenomenological
model relies on the applicability of a Ginzburg-Landau approach. Although
this seems to be the case in URu$_{2}$Si$_{2}$ given the behavior
of the thermodynamic quantities across $T_{c}$, large SC fluctuations
have been recently proposed in Ref. \cite{Yamashita15}. As for the existence
of two SC transitions, the temperature dependence of $H_{c1}$ has
been interpreted as indirect evidence for one SC transition at $T_{c}\approx1.5$
K followed by a second one at $T_{c}^{*}\approx1.2$ K \cite{Shibauchi12}. In
contrast, recent Kerr data seem to be consistent with time-reversal
symmetry being broken at $T_{c}$ \cite{Schemm14}. If this is indeed the
case, it would imply that the tetragonal symmetry breaking at the
HO transition is inconsequential for SC. On the other hand, the same
data set reveals an anomalous Kerr signal well below $T_{c}$, at
$T\approx1$ K. This anomaly, combined with a ``background'' Kerr
signal that onsets at high temperatures, indicates that at least for
now one cannot rule out the possibility of two SC transitions in URu$_{2}$Si$_{2}$.
As pointed out in Ref. {[}49{]}, additional data are needed to settle
this issue. An interesting possibility would be to perform Kerr measurements
in strained samples. Due to the dependence of $T_{c}^{*}$ on the
strain field, such a measurement would elucidate whether the anomalous
Kerr signal could be a manifestation of a second SC transition.

Our phenomenological results offer robust benchmarks
that can be employed to study the interplay between the tetragonal
symmetry breaking promoted by the HO state and the time-reversal symmetry-breaking
promoted by the SC state in URu$_{2}$Si$_{2}$. With appropriate modifications, our model
should also be relevant to other systems in which chiral SC states
have been proposed, such as the ruthenates \cite{Mackenzie14} and
doped graphene \cite{Nandkishore12}.

We thank A. Chubukov, I. Fisher, R. Flint, A. Kamenev, A. Maharaj,
Y. Matsuda, and S. Raghu for fruitful discussions. This work was supported
by the U.S. Department of Energy, Office of Science, Basic Energy
Sciences, under award number DE-SC0012336.

\end{document}